\documentclass{article}
\usepackage{graphicx}
\usepackage[english]{babel}
\usepackage{float}
\usepackage{hyperref}

\title{
\includegraphics[width=0.85\textwidth]{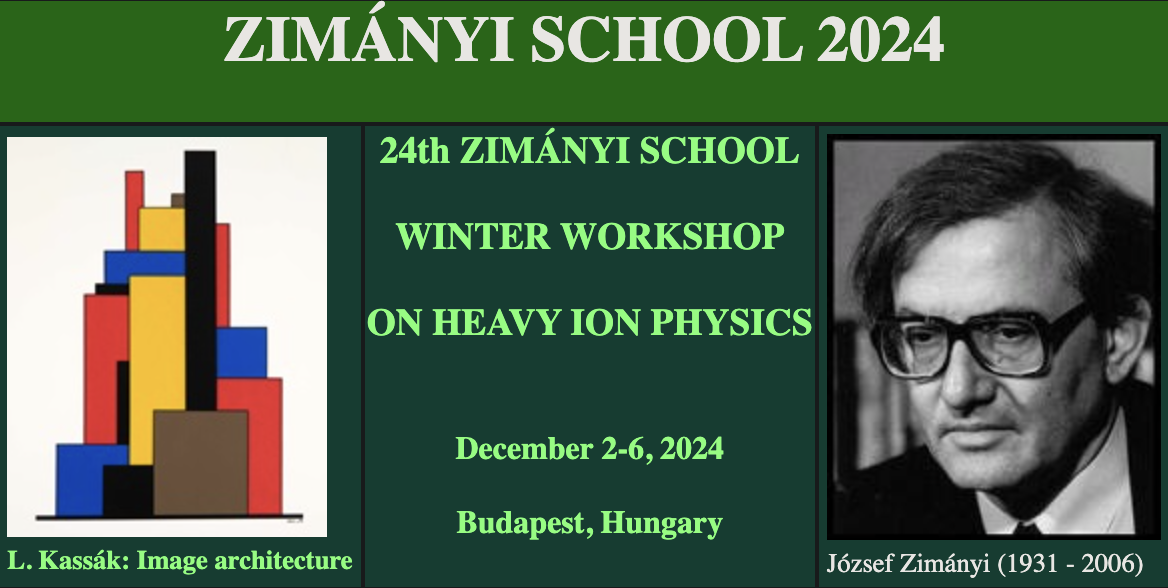}\\[1cm]
Measurements of $D^{0}$ and $D^{*}$ production in p+p collisions at $\sqrt{s}$ = 510 GeV in STAR experiment}
\author{{Subhadip Pal$^1$  (for the STAR Collaboration)}\\[1ex]
$^1$ Faculty of Nuclear Sciences and Physical Engineering, \\ Czech Technical University in Prague, \\  Břehová 7, Prague 1,  115 19, Czech Republic\\
}

\begin{document}

\fontfamily{lmss}\selectfont
\maketitle

\begin{abstract}
\noindent
These proceedings present an investigation into the production of $D^{0}$ and $D^{*}$ mesons as a function of transverse momentum in proton-proton (p+p) collisions at a center-of-mass energy of $\sqrt{s}$ = 510 GeV, in the STAR experiment at the Relativistic Heavy Ion Collider (RHIC). Objective of this analysis is to test the perturbative QCD calculations with the charm-anticharm production cross-sections obtained through $D^{0}$ and $D^{*}$ measurements. This report includes ongoing signal extractions of the $D^{0}$ and $D^{*}$ mesons from minimum bias events recorded during the p+p collisions at $\sqrt{s}$ = 510 GeV at STAR in 2017. Signals were reconstructed through the hadronic decay channels of these mesons. Like-sign combination and track rotation methods have been used to estimate the combinatorial background for $D^{0}$ measurement from $p_{T}$ = 0.0 to 2.1 GeV/$c$. For $D^{*}$, wrong-sign and side band combination of the decay daughters were utilized to reconstruct the combinatorial background from $p_{T}$ = 2.0 to 6.0 GeV/$c$. 
\end{abstract}

\section{Introduction}
\label{sec:intro}

Heavy-flavor hadrons, which consist of charm or bottom quarks, are crucial for the investigation of Quantum Chromodynamics (QCD) in high-energy hadronic collisions. Studying their production allows for comparisons between experimental results and theoretical models. Firstly, the mass of the heavy quark ($m_{Q}$) serves the purpose of a low-energy (or long distance) cut-off, thereby making it possible to calculate the processes within the regime of perturbative QCD (pQCD) up to low transverse momentum ($p_{T}$) \cite{Andronic:2016}. Also due to the occurence of multiple hard scales ($m_{Q}$, $p_{T}$), it is possible to examine the perturbation series across different kinematic regions (i.e., $p_{T} < m_{Q}$, $p_{T} \approx m_{Q}$, $p_{T} > m_{Q}$)\cite{Andronic:2016}. Based on a factorisation scheme, the differential cross section for the inclusive production of a heavy quark Q can be calculated as follows:

\begin{equation}
d\sigma^{Q+X}\simeq\sum_{i,j}f_{i}^{A}\otimes f_{j}^{B}\otimes\mathrm{d}\tilde{\sigma}_{ij\to Q+X}
\end{equation}

The parton distribution functions (PDF) $f_{i}^{A} (f_{j}^{B})$ give the number density of the parton of flavor ‘$i$’ (‘$j$’) inside the hadron ‘A’ (‘B’). d$\tilde{\sigma}$ is the partonic cross-section which depends on the strong-coupling constant ($\alpha_{s}$) \cite{Andronic:2016}. One can subsequently convolute this differential cross section for the production of the heavy quark Q with a suitable, scale-independent, fragmentation function $D^{H}_{Q} (z)$ which describes the transition of the heavy quark with momentum $p_{Q}$ into the observed heavy-flavored hadron $H$ with momentum $p_{H} = z p_{Q}$ \cite{Andronic:2016}: 

\begin{equation}
d\sigma^{H+X} = d\sigma^{Q+X} \otimes D_Q^H(z).
\end{equation}

 The pQCD calculations of Fixed-Order Next-to-Leading Logarithm (FONLL)  can describe the heavy quark production at transverse momenta $p_{T}$ larger than their mass \cite{Nelson:2013}. But, in the low momentum region (e.g., $p_{T} < 1$ GeV/$c$ for charm quarks) the strong coupling constant increases drastically where pQCD has restricted ability to predict the cross-section. Precise experimental measurements become necessary to constrain calculations and improve the low-$p_{T}$ modeling.

\begin{figure}[ht]
  \centering
    \includegraphics[width=0.6\textwidth]{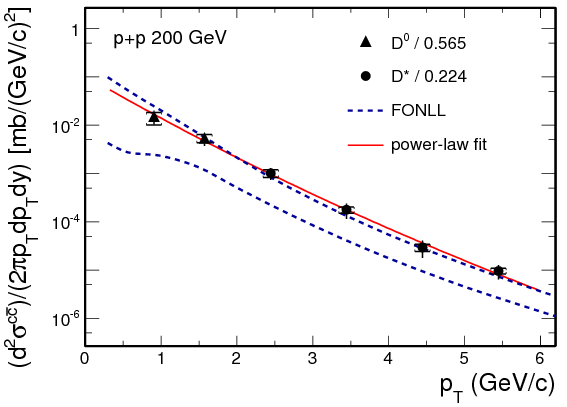}
    \caption[$p_T$-differential $c\bar{c}$ production cross section in p+p collisions at $\sqrt{s}$ = 200 GeV]{$p_T$-differential $c\bar{c}$ production cross section measured in p+p collisions at $\sqrt{s}=200$ GeV compared with Fixed Order Next-to-Leading Logarithm (FONLL) pQCD calculations. The $D^0$ and $D^{*}$ data points were divided by the corresponding charm quark fragmentation ratios 0.565 $\left(c \rightarrow D^0\right)$ and 0.224 $\left(c \rightarrow D^*\right)$ \cite{Adamczyk:2012}.}
    \label{fig:DmesonSTAR2012}
\end{figure}

The investigation of $D$ meson production in proton+proton (p+p) collisions at the Relativistic Heavy Ion Collider (RHIC) and the Large Hadron Collider (LHC) has attracted considerable interest, as it holds important implications for the comprehension of QCD and the mechanisms underlying charm quark production \cite{Adamczyk:2012, Acharya:2017, Acharya:2019}. The STAR collaboration has reported measurements of $D$ mesons in p+p collisions at $\sqrt{s}$ = 200 GeV. The $p_T$-differential $c\bar{c}$ production cross section from $D^0$ and $D^{*}$ measurements was compared with FONLL pQCD calculations, as illustrated in Fig.~\ref{fig:DmesonSTAR2012} \cite{Adamczyk:2012}. 

\begin{figure}[ht]
  \centering
    \includegraphics[width=0.7\textwidth]{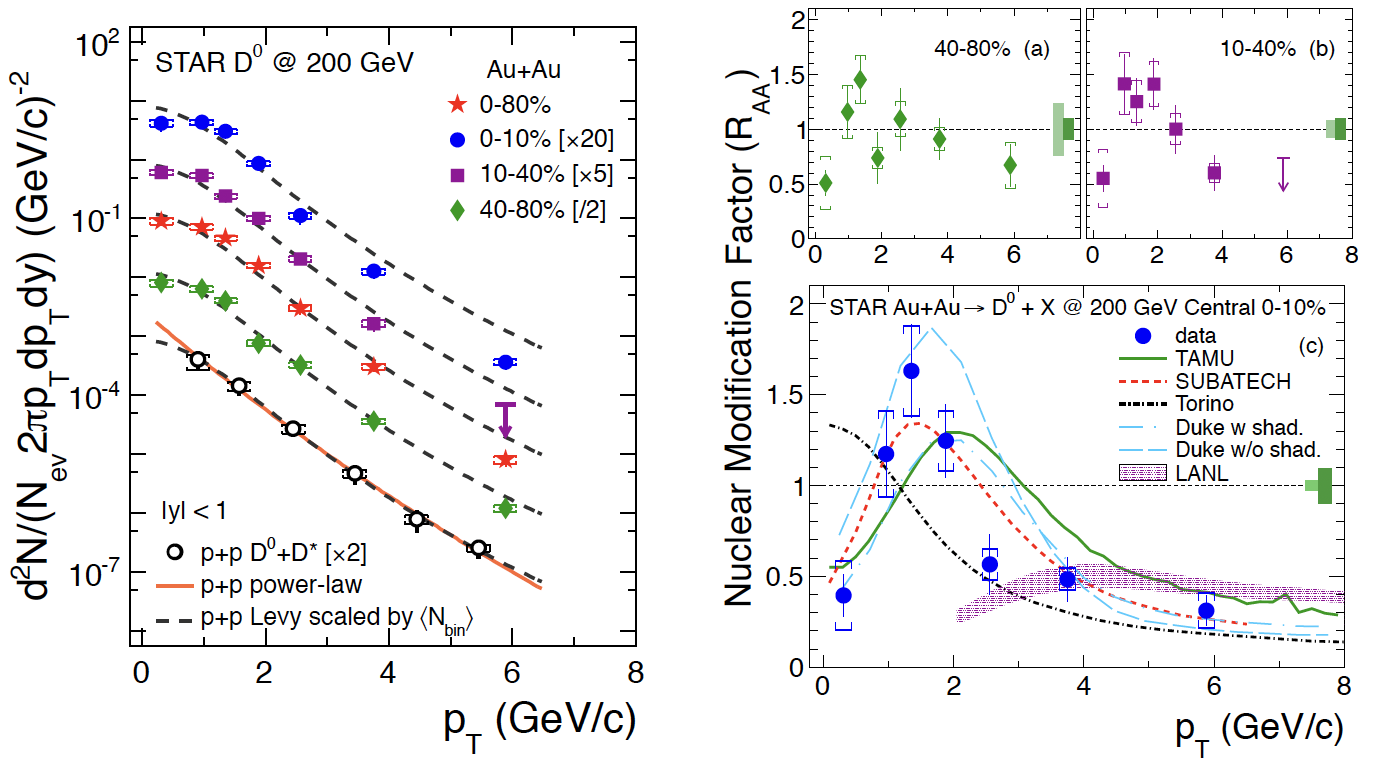}
    \caption[D$^{0}$ meson $R_{AA}$ in Au+Au collisions at $\sqrt{s_{NN}}$ = 200 GeV]{$p_T$-differential $D^0$ invariant yield in Au+Au collisions at $\sqrt{s_{NN}}=200$ GeV for various centralities (left) and  the $D^{0}$ meson nuclear modification factor $R_{AA}$ (right) \cite{Adamczyk:2014}.}
    \label{fig:DmesonAuAu}
\end{figure}

In addition, modifications of the charm meson production in heavy-ion collisions with respect to p+p provide insights into the presence of Quark Gluon Plasma (QGP) medium. The $D^{0}$ meson nuclear modification factor ($R_{AA}$) in Au+Au collisions at a center-of-mass energy $\sqrt{s_{NN}}=200$ GeV at RHIC has been studied previously to understand the suppression effects due to the presence of the QGP \cite{Nahrgang:2014, Xie:2016}.   The results indicate a strong suppression of $D^{0}$ yields at high transverse momentum, which is a signature of the medium's influence on heavy quark dynamics. The $p_T$-differential invariant yield of $D^0$ in Au+Au collisions at $\sqrt{s_{NN}}=200$ GeV, across various centralities, along with the nuclear modification factor for $D^{0}$, are presented in Fig.~\ref{fig:DmesonAuAu}.

\section{Measurement of Open-Charm Mesons \\ with STAR} 
\label{sec:measurement-star}

In 2017, STAR collected about 1.11 billion minimum-bias events for p+p collisions at $\sqrt{s}$ = 510 GeV which allows precise heavy flavor meson measurements. In this section, the methodology of the signal extraction of open-charm mesons $D^{0}$ and $D^{*}$ from this data is described. 

As $D^{0}$ and $D^{*}$ mesons decay before reaching the detectors, it is not possible to directly observe those. Therefore, it is necessary to reconstruct those through their decay products. In this analysis, the following decay channels have been utilized:  $D^0 (\bar{D^0}) \rightarrow K^\mp \pi^\pm (\mathrm{Branching \ Ratio} = 3.947\pm0.030 \%)$ and  $D^{*\pm} \rightarrow D^{0} (\bar{D^0}) \pi^{\pm}_{s}  (\mathrm{Branching \ Ratio} = 67.7\pm0.5 \%) \rightarrow K^{\mp} \pi^{\pm} \pi^{\pm}_{s}$ \cite{PDG:2024}. Here, $\pi_{s}$ denotes the soft pions directly decaying from $D^{*}$.

A set of event selection cuts were applied on this minimum bias triggered data. The events were selected so that the position of the primary vertex along the beam axis $\mathrm{V_{z [TPC]}}$ determined using the Time Projection Chamber (TPC) \cite{Anderson:2003} is no further than 60 cm from the center of STAR. A cut on the difference of vertex z coordinate measured by TPC ($\mathrm{V_{z [TPC]}}$) and by Vertex Position Detector \cite{Llope:2014} ($\mathrm{V_{z [VPD]}}$) was imposed to remove pile-up vertices. Also, only primary vertices close to the beamline were selected through the cuts on  $\mathrm{V_{x [TPC]}}$ and  $\mathrm{V_{y [TPC]}}$.

Tracks are reconstructed in TPC from registered hits - therefore the higher the number of hits, the more precise is the final track \cite{Anderson:2003}. Only tracks reconstructed from more than 18 points out of maximum possible 45 TPC fit points were kept in this analysis.  The ratio of the number of TPC fit points to the maximum possible number of TPC fit points, exceeding 0.52, was employed to reduce the contribution from split tracks, wherein a single track is erroneously reconstructed as two distinct tracks. A 1.5 cm limit on DCA (Distance of Closest Approach of a track to the primary vertex) was found to be effective on removing pile-up tracks. Particles were needed to have a certain minimal value of transverse momentum ($p_{T}$) to be accepted for the sake of higher reconstruction efficiency. Also, tracks within the acceptance of the TPC and Time of Flight (TOF) detectors \cite{Shao:2006} were selected by imposing a cut on the pseudorapidity $|\eta|$. An overview of these track quality cuts and event selection cuts have been demonstrated in Table~\ref{tab:eventtrackcut}.

\begin{table}[ht]
\begin{center}
\footnotesize
\caption{Summary of event and track selection criteria used in this analysis.}
\begin{tabular}{cc}
\hline
\hline
& $\left|\mathrm{V_{z [TPC]}}-\mathrm{V_{z [VPD]}}\right|$  $<4.0 \mathrm{~cm}$ \\
Event Selection Cuts & $\mathrm{~V_{z [TPC]}}$  $<60 \mathrm{~cm}$\\
& $\mathrm{~V_{x [TPC]}} $   $ \in (-0.3, 0.14) \mathrm{~cm} $ \\
& $\mathrm{~V_{y [TPC]}} $   $ \in (-0.26, 0.02) \mathrm{~cm}$ \\
\hline
\hline
& number of TPC fit points   $> 18$ \\
& $\frac{\mathrm{number \ of \ TPC \ fit \ points}}{\mathrm{number \ of \ max \ possible \ TPC \ fit \ points}}$  $ > 0.52 $\\
Track Quality Cuts & global DCA   $ < 1.5 \mathrm{~cm} $ \\
& $p_{T}   > 0.2$ GeV/$c$ \textsuperscript{1} \\
& $|\eta|$  $ < 1$ \\
\hline
\hline

\end{tabular}
 \label{tab:eventtrackcut}
\end{center}
\noindent{\footnotesize{\textsuperscript{1} To select $\pi_{s}$ candidates directly decaying from $D^{*}$, $p_{T}$ cut was reduced to 0.1 GeV/$c$ to select lower momentum tracks.}}
\end{table}

Next, the daughter particle candidates were identified by utilizing the TPC and TOF detectors. The identification of charged particles is achieved through the measurement of energy loss within the TPC gas. The measured energy loss is compared to the expected one. Particle identification in TOF is done by comparing measured value of inverse velocity $1/\beta_\mathrm{TOF}$ with the expected theoretical value $1/\beta_\mathrm{th}$. $\beta_\mathrm{th}$ is obtained from the particle momentum $p$ measured by TPC and expected rest mass $m$ using the formula $\beta_\mathrm{th}=p/\sqrt{p^2+m^2c^2}$ \cite{Shao:2006}. 

In order to enhance observed signals, $D^{0}$ and $\bar{D^{0}}$ were analyzed together. Unlike-sign pion and kaon candidates were paired and the invariant mass of the resulting pair was computed. Pion-kaon pairs with a rapidity outside the interval of (-1, 1) were excluded from consideration. As most of the $K\pi$ pairs do not come from $D^0 (\bar{D^{0}})$  decay, one has to deal with a huge combinatorial background. The combinatorial background was reconstructed using two independent methodologies: (i) like-sign combination where candidates for pions were paired with kaon candidates of identical charge and (ii) track rotation where one of the tracks was rotated so that its momentum was pointing in opposite direction: $(E,p_x,p_y,p_z)\longrightarrow(E,-p_x,-p_y,-p_z)$ , thus destroying all correlation and decay kinematics. 

Now, in order to form a $D^{*}$ candidate, a soft pion candidate ($\pi_{s}$) was combined with the unlike-sign $K\pi$ pair if the invariant mass of the pair $M_{K^\mp\pi^\pm}$ falls between 1.8 and 1.95 GeV/$c^{2}$. The invariant mass of the $K^\mp\pi^\pm\pi^\pm_s$ triplet $M_{K^\mp\pi^\pm\pi^\pm_s}$ was subsequently calculated, and a histogram was populated with the mass difference $M_{K^\mp\pi^\pm\pi^\pm_s}$ - $M_{K^\mp\pi^\pm}$. Furthermore, the combinatorial background was reconstructed utilizing two distinct and independent methods: wrong-sign combination method and side-band method. In the wrong-sign combination, the soft pion was paired with the $D^{0}$ daughter pion of the opposite charge. In the side-band method, the $M_{K^\mp\pi^\pm}$ had been lying between two side-bands: 1.64 - 1.74 GeV/$c^{2}$ and 2.01 - 2.11 GeV/$c^{2}$, i.e., outside the $D^{0}$ mass window.

\section{Results}

\begin{figure}[ht]
\centering
\begin{minipage}{0.32\textwidth}
      \centering
      \includegraphics[width=\textwidth]{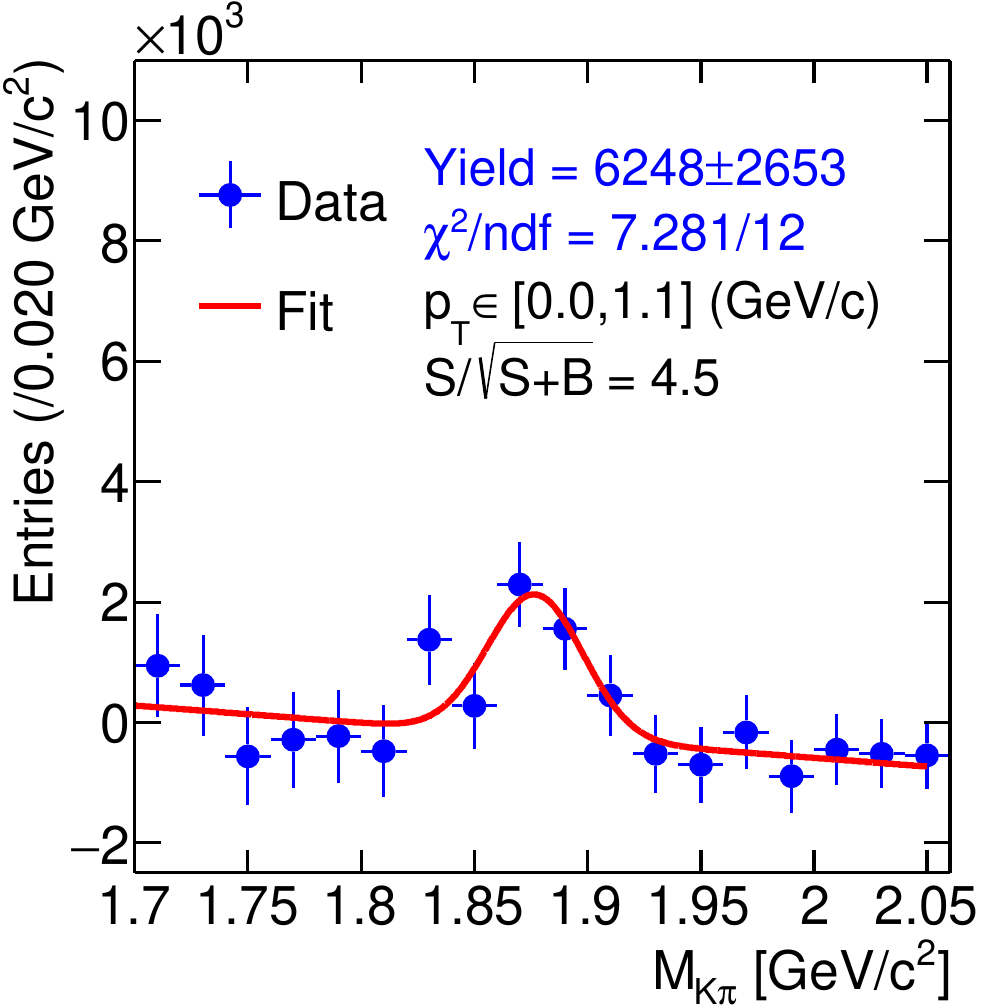}
\end{minipage}
\hfill
\begin{minipage}{0.32\textwidth}
      \centering
      \includegraphics[width=\textwidth]{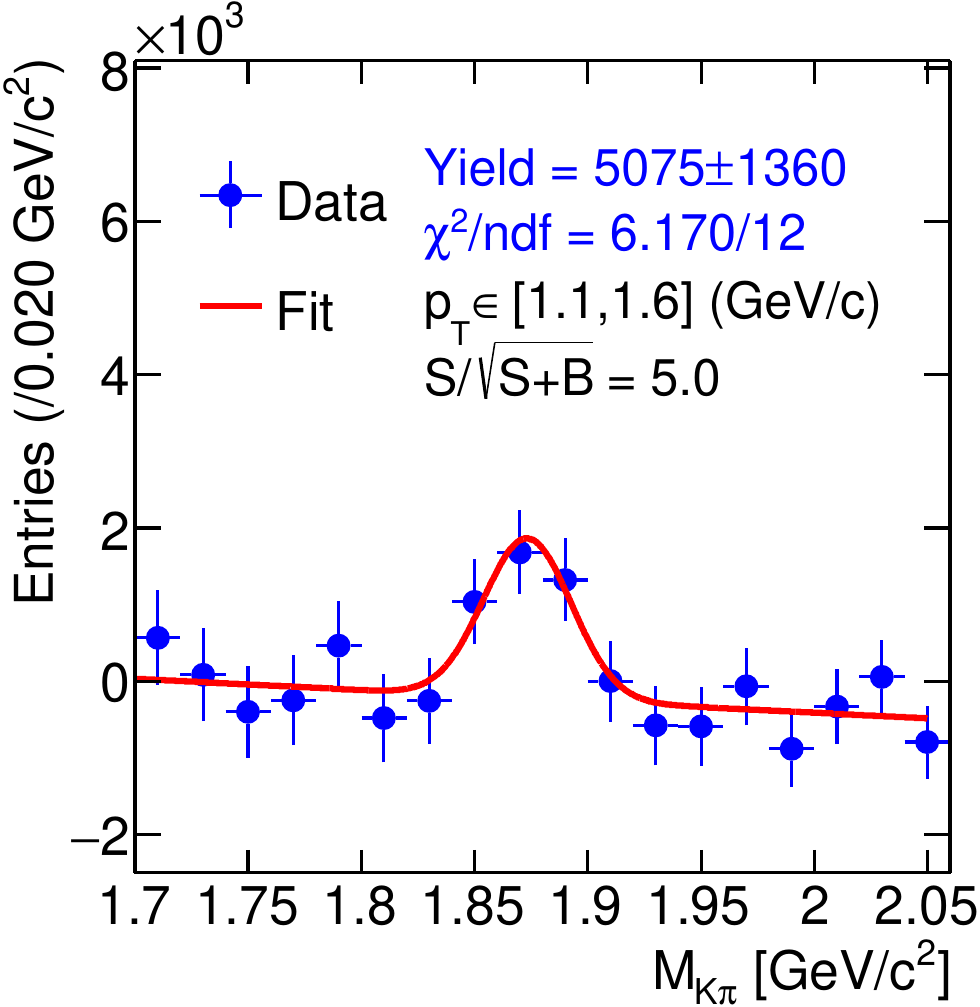}
\end{minipage}
\hfill
\begin{minipage}{0.32\textwidth}
      \centering
      \includegraphics[width=\textwidth]{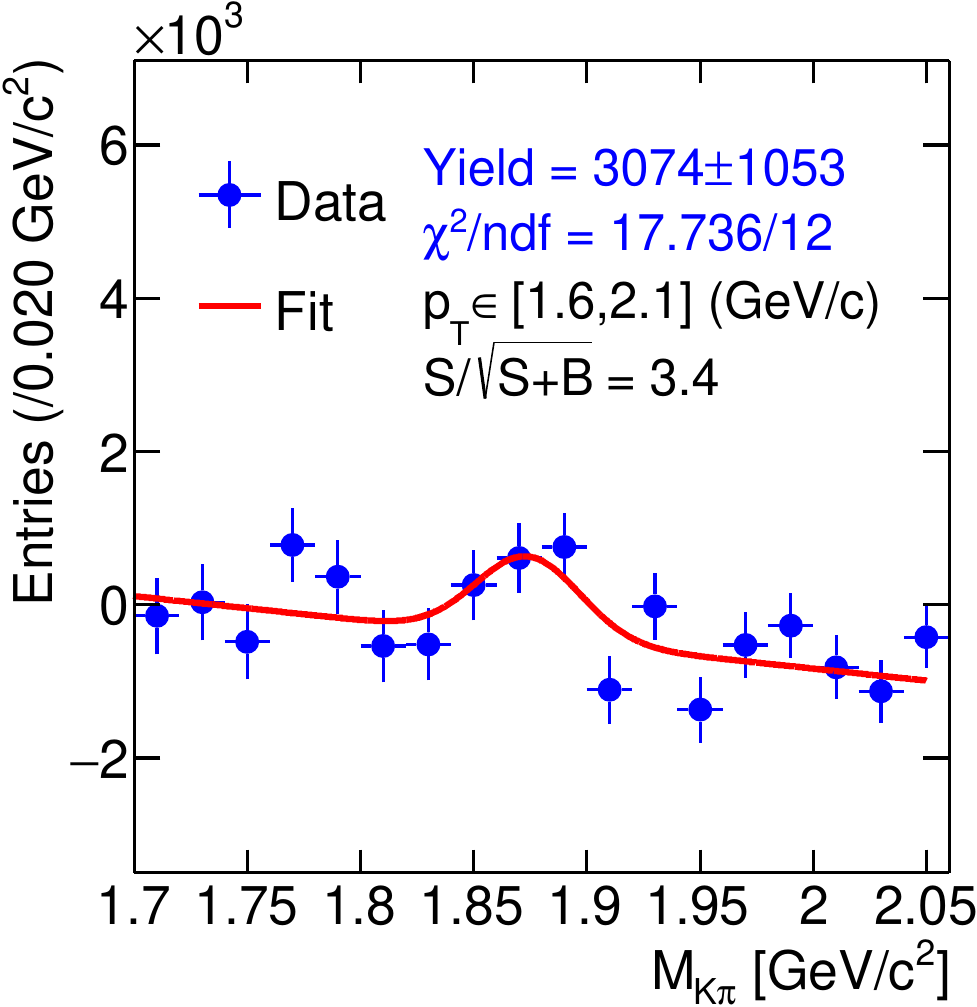}
\end{minipage}
\caption{Like-sign background subtracted invariant mass distributions of $D^0$ candidates in three $p_{T}$ bins fitted with Gaussian+linear function.}
\label{fig:D0_mass_ptbins}
\end{figure}

The $D^0$ signals were extracted by fitting the combinatorial background subtracted invariant mass distribution with a sum of Gaussian and linear functions. Mean, standard deviation and the area under the curve were kept to be free parameters for the Gaussian function. The raw yield was determined by calculating the area under the Gaussian distribution superimposed on a linear function, which was employed to accurately model the residual background with sufficient precision. Like-sign background subtracted mass distribution of $D^0$ candidates with 0.0 $\leq$ $p_{T}$ $\leq$ 1.1 GeV/$c$, 1.1 $\leq$ $p_{T}$ $\leq$ 1.6 GeV/$c$ and 1.6 $\leq$ $p_{T}$ $\leq$ 2.1 GeV/$c$ are shown in Fig.~\ref{fig:D0_mass_ptbins}.

\begin{figure}[ht]
\centering
\begin{minipage}{0.32\textwidth}
      \centering
      \includegraphics[width=\textwidth]{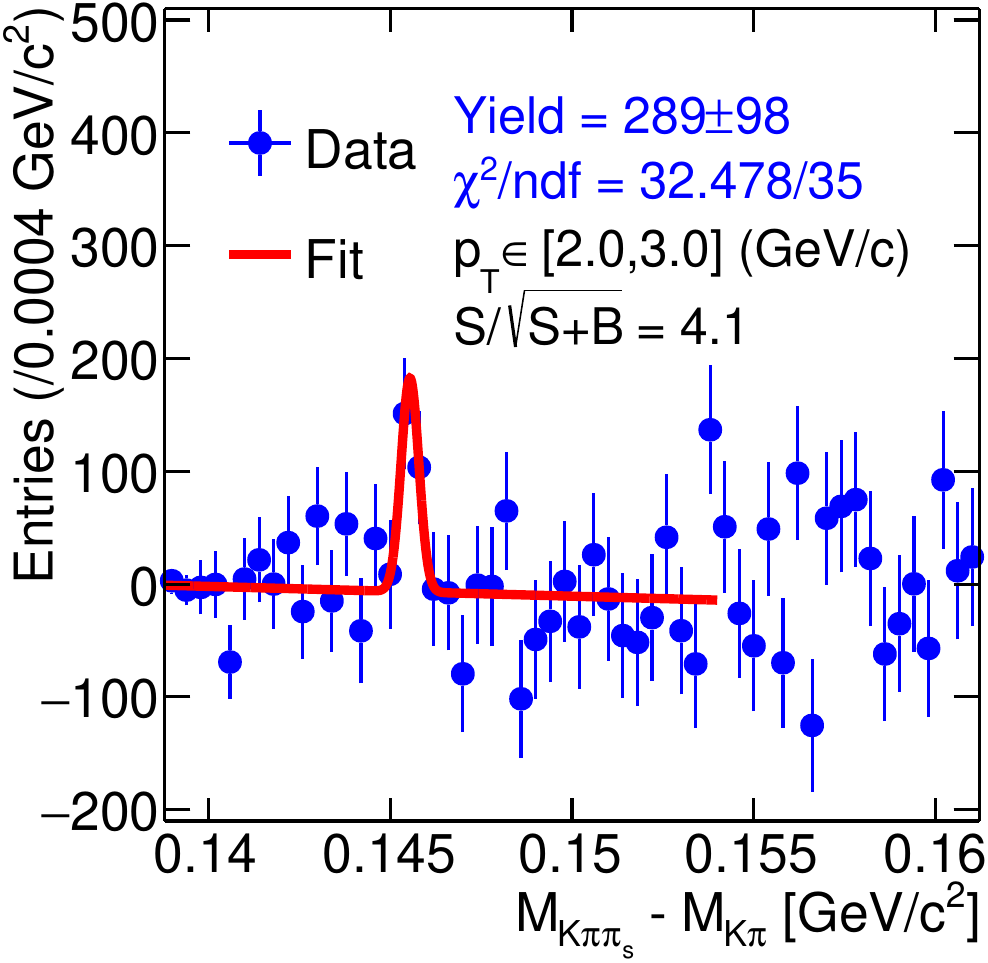}
\end{minipage}
\hfill
\begin{minipage}{0.32\textwidth}
      \centering
      \includegraphics[width=\textwidth]{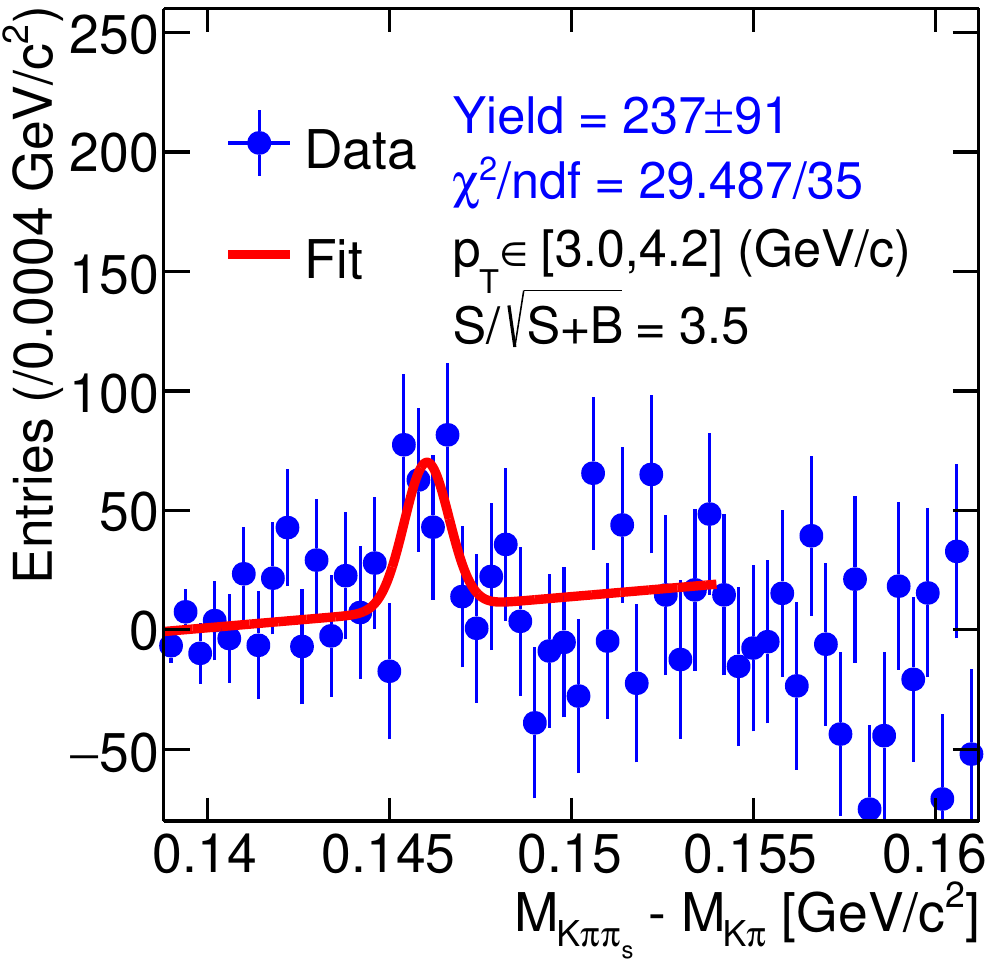}
\end{minipage}
\hfill
\begin{minipage}{0.32\textwidth}
      \centering
      \includegraphics[width=\textwidth]{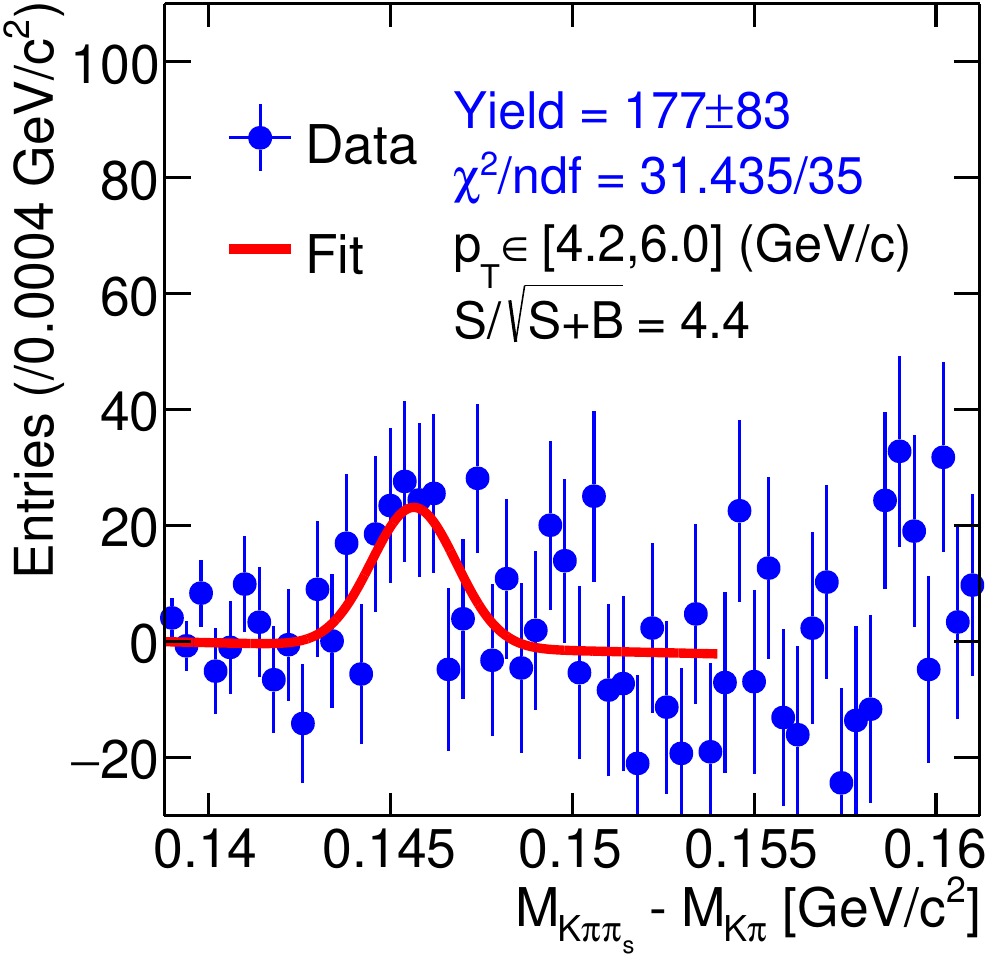}
\end{minipage}
\caption{Wrong-sign background subtracted invariant mass distributions of $D^*$ candidates in three $p_{T}$ bins fitted with Gaussian+linear function.}
\label{fig:Dstar_mass_ptbins_ws}
\end{figure}

\begin{figure}[H]
\centering
\begin{minipage}{0.32\textwidth}
      \centering
      \includegraphics[width=\textwidth]{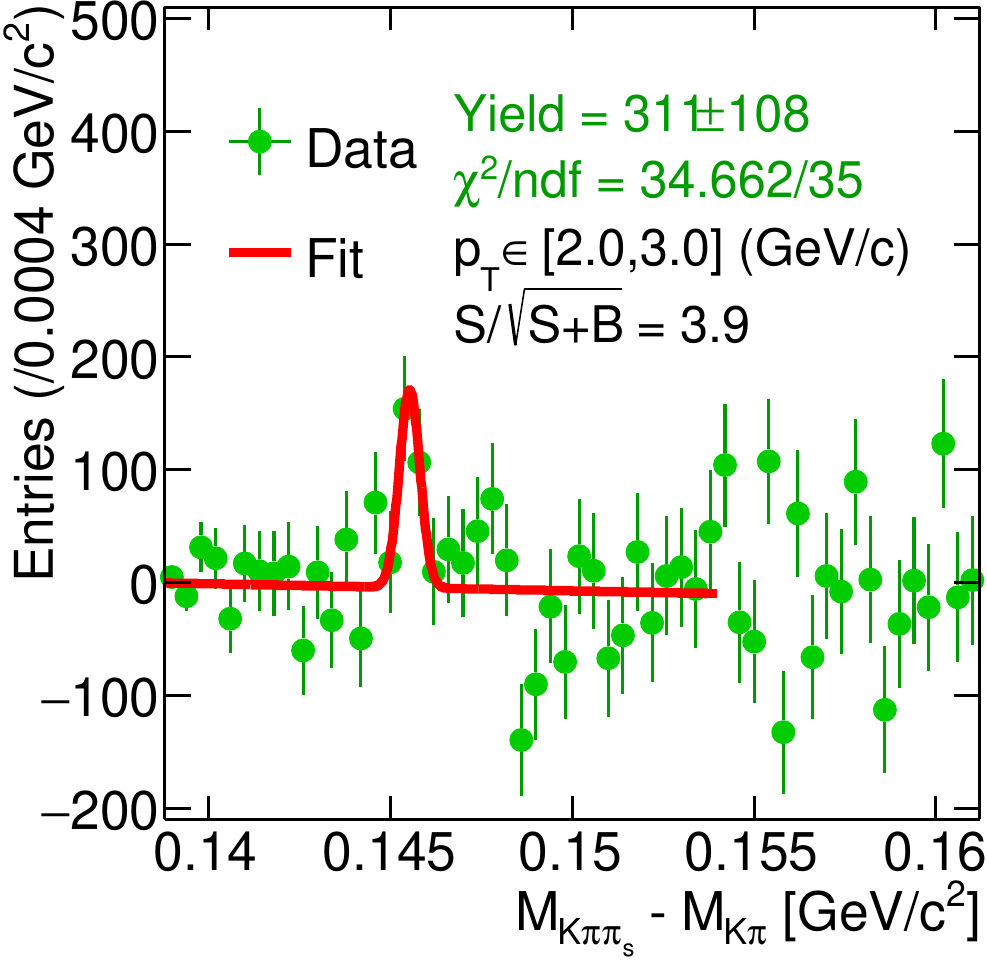}
\end{minipage}
\hfill
\begin{minipage}{0.32\textwidth}
      \centering
      \includegraphics[width=\textwidth]{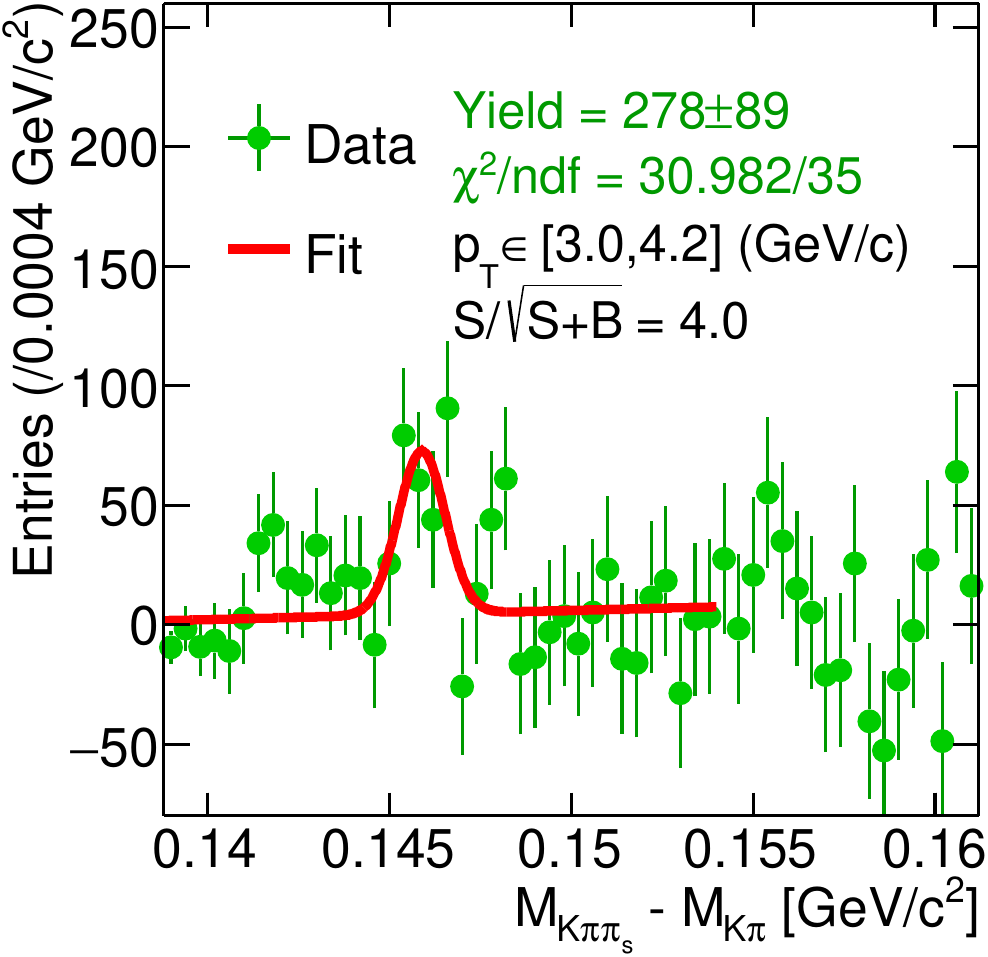}
\end{minipage}
\hfill
\begin{minipage}{0.32\textwidth}
      \centering
      \includegraphics[width=\textwidth]{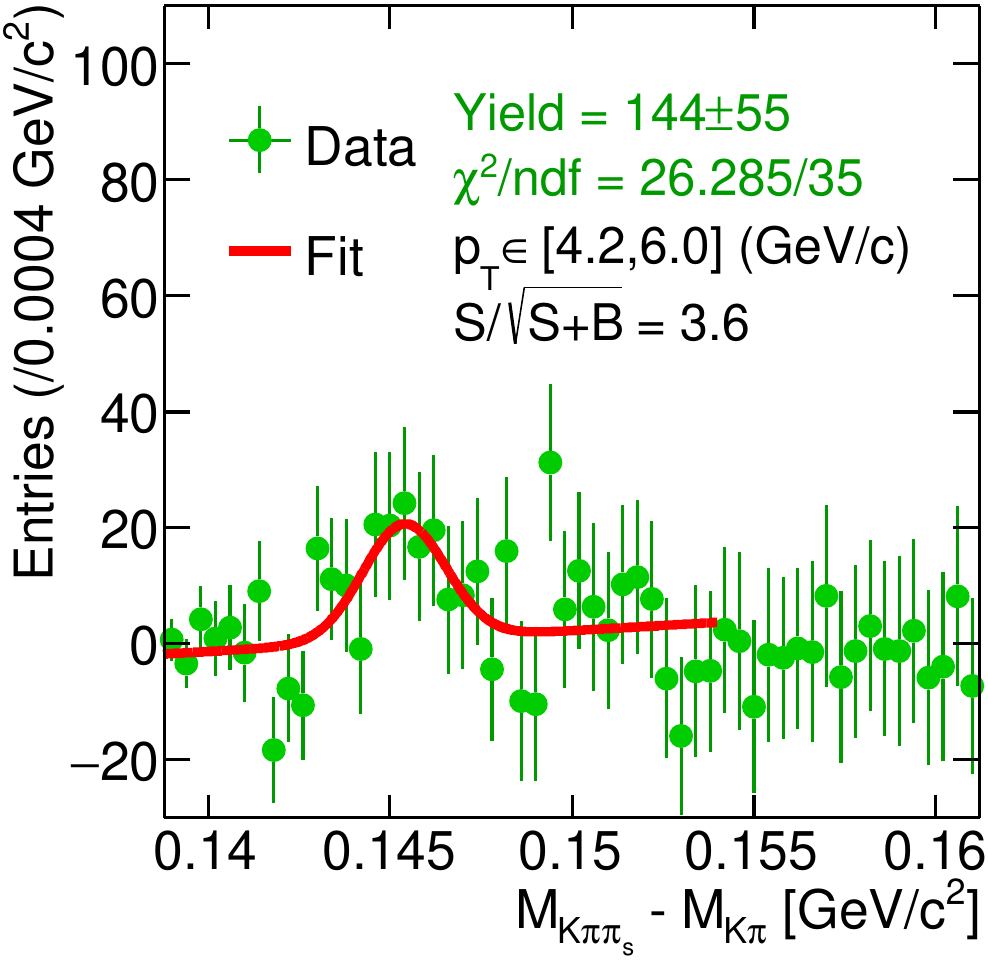}
\end{minipage}
\caption{Side-band background subtracted invariant mass distributions of $D^*$ candidates in three $p_{T}$ bins fitted with Gaussian+linear function.}
\label{fig:Dstar_mass_ptbins_sb}
\end{figure}

The $D^{*}$ signals after subtracting the combinatorial background reconstructed by the wrong-sign pairs and side-band methods in the $p_{T}$ bins 2.0 - 3.0 GeV/$c$, 3.0 - 4.2 GeV/$c$, 4.2 - 6.0 GeV/$c$ are shown in Fig.~\ref{fig:Dstar_mass_ptbins_ws} and Fig.~\ref{fig:Dstar_mass_ptbins_sb} respectively. These signals were also extracted by fitting the background subtracted invariant mass distribution with a sum of Gaussian and linear functions for each $p_{T}$ bin. Again, the raw yields were estimated from the area under the Gaussian distribution which had the mean, standard deviation and the area under the curve as free parameters. As presented in Fig.~\ref{fig:Dstar_mass_ptbins_ws} and Fig.~\ref{fig:Dstar_mass_ptbins_sb}, the extracted $D^{*}$ raw yields from the two independent background reconstruction methods (wrong-sign pairs and side-band) were consistent within uncertainties.

The significance of both of the $D^{0}$ and $D^{*}$ signals were calculated as $S/\sqrt{S+B}$, where $S$ is the raw yield of the signal and $B$ is the raw yield of the background.

\section{Discussion} 
\label{sec:discussion}

In this analysis, $D^{0}$ and $D^{*}$ measurements in p+p collisions at $\sqrt{s} = 510$ GeV has been presented. These open-charm mesons were reconstructed through their hadronic decay channels at STAR.  $D^{0}$ candidates were reconstructed from $p_{T}$ = 0.0 to 2.1 GeV/$c$. Wrong-sign and Side-band methods were used to estimate the combinatorial background of the $D^{*}$ candidates' invariant mass spectrum from $p_{T}$ = 2.0 to 6.0 GeV/$c$. Since, TOF had been used here to identify the soft pions from $D^{*\pm} \rightarrow D^{0} (\bar{D^0}) \pi^{\pm}_{s}$ decay (minimum $p_{T}$ of a track to reach TOF $\approx$ 0.16 GeV/c), it was not kinematically favorable to extract $D^{*}$ signals with lower $p_{T}$ \cite{Tlusty:2014}. Nevertheless, both $D^{0}$ and $D^{*}$ signals were extracted with a significance more than 3$\sigma$. 

In $D^{0}$ reconstruction, daughter particles might be misidentified. Kaon is then labeled as pion and vice-versa. In such case, the pair will contribute to both $D^{0}$ and $\bar{D^{0}}$ signal and cause a double-counting in the raw yield estimation. For the reconstruction of $D^*$ mesons via the decay channel $D^{*\pm} \rightarrow D^0\pi_S^\pm \rightarrow K^\mp\pi^\pm\pi_S^\pm$, a distinct scenario emerges. In the event that $K^\mp$ is mistakenly identified as $\pi^{\mp}$ and $\pi^\pm$ as $K^\pm$, the resulting combination $K^\pm\pi^\mp\pi_S^\pm$ fails to contribute to the $\Delta M$ peak, instead augmenting the wrong-sign background. Consequently, the raw yield obtained through the subtraction of the wrong-sign background will be undercounted. The probability of such double-counting (undercounting) in $D^{0}$ ($D^{*}$) measurements can be obtained from Monte Carlo simulations.

The primary objective of this analysis is to measure the charm production cross-section in proton-proton collisions at a center-of-mass energy of $\sqrt{s}$ = 510 GeV. To obtain that one also needs to take into account the efficiency of the experiment, trigger biases, and systematic uncertainties. The measured transverse momentum differential invariant charm production cross section obtained from the $D^{0}$ and $D^{*}$ measurements will be subjected to a fit to the Lévy power-law function \cite{Wong:2013} in order to extrapolate the cross section outside of the measured kinematic region and it would be subsequently compared to the predictions derived from the FONLL framework.

\section*{Acknowledgments} 
\label{sec:acknowledgement}

The work was supported by the Grant Agency of the Czech Technical University in Prague, grant No. SGS25/159/OHK4/3T/14 and by the Czech Science Foundation, grant no. GA23-07499S as well as partial support from the U.S. Department of Energy (DOE).

\end{document}